\documentclass[twocolumn,notitlepage,nofootinbib]{revtex4-2}
\usepackage{amsmath,amssymb,bm,graphicx,hyperref}
\allowdisplaybreaks[1]

\renewcommand\d{\partial}
\newcommand\+{\dagger}
\newcommand\<{\langle}
\renewcommand\>{\rangle}
\newcommand\eps{\varepsilon}
\newcommand\ep{\eps_p}
\newcommand\eq{\eps_q}
\newcommand\eqq{\eps_{q'}}
\newcommand\R{\mathbb{R}}
\newcommand\Z{\mathbb{Z}}
\newcommand\C{\mathcal{C}}
\newcommand\E{\mathcal{E}}
\renewcommand\H{\mathcal{H}}
\newcommand\N{\mathcal{N}}
\renewcommand\S{\mathcal{S}}
\newcommand\pv{\mathrm{P.V.}}
\newcommand\reg{\mathrm{reg}}

\let\Im\relax\DeclareMathOperator\Im{Im}

\begin{document}

\title{Viscous Drude weight of dual Bose and Fermi gases in one dimension}

\author{Yusuke Nishida}
\affiliation{Department of Physics, Tokyo Institute of Technology,
Ookayama, Meguro, Tokyo 152-8551, Japan}

\date{October 2022}

\begin{abstract}
We continue to study frequency-dependent complex bulk viscosities of one-dimensional Bose and Fermi gases with contact interactions, which exhibit the weak-strong duality according to our recent work.
Here we show that they are contributed to by Drude peaks divergent at zero frequency as typical for transport coefficients of quantum integrable systems in one dimension.
In particular, their Drude weights are evaluated based on the Kubo formula in the high-temperature limit at arbitrary coupling as well as in the weak-coupling and strong-coupling limits at arbitrary temperature, where systematic expansions in terms of small parameters are available.
In all three limits, the Drude peaks are found at higher orders compared to the finite regular parts.
\end{abstract}

\maketitle

\section{Introduction}
A nonvanishing Drude weight indicates a divergent transport coefficient at zero frequency and serves as diagnostics of whether the transport is ballistic or diffusive~\cite{Kohn:1964}.
A simple example of ballistic transports is provided by a mass transport for fluids with translational invariance, where the mass current (i.e., momentum) cannot dissipate due to its conservation law~\cite{Mahan}.
On the other hand, an energy transport is typically diffusive for interacting systems because the energy current is nonconserved.
However, quantum integrable systems in one dimension have been found so exceptional that their Drude weights remain nonvanishing for various transports even when corresponding currents are nonconserved~\cite{Castella:1995,Zotos:1997}.
This is because a macroscopic number of conservation laws allows the currents to have some overlaps with conserved quantities.
Since then, anomalous conductivities of quantum integrable systems in one dimension have been subjected to active study from both theoretical and experimental perspectives~\cite{Zotos:2004,Zotos:2005,Sirker:2020,Bertini:2021}.

In spite of such active study, little attention has been paid so far to another transport coefficient possible in one dimension, that is, the bulk viscosity~\cite{Matveev:2017,DeGottardi:2020}.
Recently, we showed in Ref.~\cite{Tanaka:bulk} that the frequency-dependent complex bulk viscosity of a Bose gas with a contact interaction known as the Lieb-Liniger model~\cite{Lieb:1963a,Lieb:1963b} is identical to that of a dual Fermi gas known as the Cheon-Shigehara model~\cite{Cheon:1999} at the same scattering length $a$.
In particular, it is the weak-strong duality, where one system at weak coupling corresponds to the other system at strong coupling so that the bulk viscosity in the strong-coupling regime can be accessed with the perturbation theory of the dual system.
Their bulk viscosities were then computed in the high-temperature, weak-coupling, and strong-coupling limits, where systematic expansions in terms of small parameters are available, and found to be finite with no Drude peaks at their leading orders~\cite{Tanaka:bulk}.

\begin{figure}[b]
\includegraphics[width=\columnwidth]{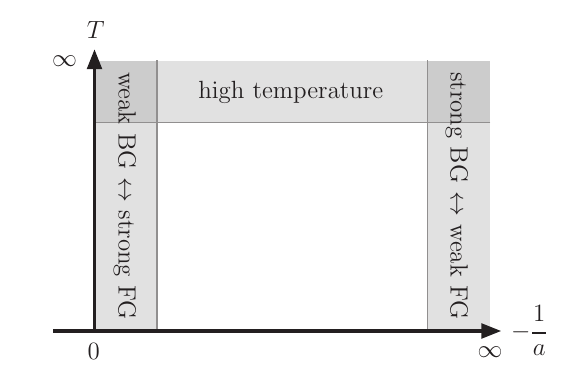}
\caption{\label{fig:duality}
Bulk viscosities of Bose and Fermi gases are evaluated in the high-temperature limit as well as in the weak-coupling limit, which corresponds to $a\to-\infty$ for bosons (BG) and $a\to-0$ for fermions (FG)~\cite{Tanaka:bulk}.
The system is thermodynamically unstable at $a>0$.}
\end{figure}

The purpose of this paper is to go one step further beyond the leading orders and show that the frequency-dependent complex bulk viscosities of one-dimensional Bose and Fermi gases with contact interactions indeed have the structure of
\begin{align}\label{eq:peak}
\zeta(\omega) = \zeta_\reg(\omega) + \frac{iD}{\omega+i0^+},
\end{align}
where the first term on the right-hand side is the finite regular part and the second term is the Drude peak divergent at zero frequency.
In particular, we will evaluate the Drude weights $D$ based on the Kubo formula in the high-temperature limit at arbitrary coupling as well as in the weak-coupling limit at arbitrary temperature for bosons in Sec.~\ref{sec:bose} and for fermions in Sec.~\ref{sec:fermi}.
The two results in the high-temperature limit are also useful to confirm the Bose-Fermi duality explicitly, whereas those in the weak-coupling limit are applicable to the Fermi and Bose gases in the strong-coupling limit, as indicated in Fig.~\ref{fig:duality}.

We will set $\hbar=k_B=1$ throughout this paper and the bosonic and fermionic frequencies in the Matsubara formalism are denoted by $p_0=2\pi n/\beta$ and $p'_0=2\pi(n+1/2)/\beta$, respectively, for $n\in\Z$ and $\beta=1/T$.
Also, an integration over wave number or momentum is denoted by $\int_p\equiv\int_{-\infty}^\infty dp/2\pi$ for the sake of brevity, whereas the same definition as in Ref.~\cite{Fujii:2020} is employed for the response function [see Eq.~(10) therein].

\section{Drude weight}
According to the linear-response theory~\cite{Mori:1962,Luttinger:1964,Bradlyn:2012}, the complex bulk viscosity at frequency $\omega$ is microscopically provided by
\begin{align}\label{eq:kubo}
\zeta(\omega) = \frac{R_{\Pi\Pi}(w) - R_{\Pi\Pi}(i0^+)}{iw}
\end{align}
with the substitution of $w\in\mathbb{C}\to\omega+i0^+$ on the right-hand side~\cite{Fujii:2020}.
Here $R_{\Pi\Pi}(w)$ is the response function of the modified stress operator at zero wave number,
\begin{align}\label{eq:stress}
\hat\Pi \equiv \hat\pi - \left(\frac{\d p}{\d\N}\right)_\E\hat\N
- \left(\frac{\d p}{\d\E}\right)_\N\hat\H,
\end{align}
where $\hat\pi$, $\hat\N$, and $\hat\H$ are the stress operator, the number density operator, and the Hamiltonian density, respectively, with $p=\<\hat\pi\>$, $\N=\<\hat\N\>$, and $\E=\<\hat\H\>$.
Although the above Kubo formula may look different from that employed in Ref.~\cite{Tanaka:bulk}, they are actually equivalent but the present form will prove to be convenient for the sake of evaluating the Drude weight.

\begin{figure}[t]
\includegraphics[width=0.4\columnwidth]{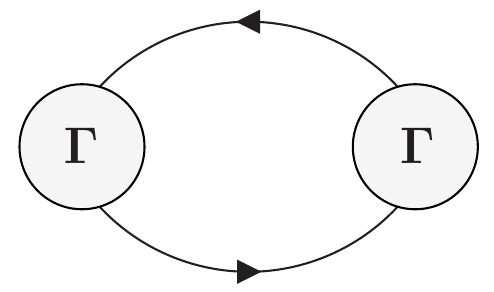}
\caption{\label{fig:drude}
Diagrammatic representation of the stress-stress response function in Eq.~(\ref{eq:drude}).
The single line represents the single-particle propagator, whereas the circle ($\Gamma$) is the vertex function.}
\end{figure}

The Drude peak appears from the type of diagrams depicted in Fig.~\ref{fig:drude} for the stress-stress response function, which reads
\begin{align}\label{eq:drude}
R_{\Pi\Pi}(ik_0) &= \pm\frac1\beta\sum_{p_0}\int_p\,G(ik_0+ip_0,p)G(ip_0,p) \notag\\
&\quad \times [\Gamma(ik_0+ip_0,ip_0;p)]^2.
\end{align}
Here $G(ip_0,p)=1/(ip_0-\ep)$ with $\ep=p^2/2m-\mu$ is the single-particle propagator, $\Gamma(ik_0+ip_0,ip_0;p)$ is the vertex function to be specified below, and the upper (lower) sign corresponds to bosons (fermions under $p_0\to p'_0$).
The Matsubara frequency summation is replaced with the complex contour integration over $ip_0\to\nu$ and its integration contour is deformed into four lines along $\Im(\nu)=\pm0^+,-k_0\pm0^+$~\cite{Fujii:2021,Tanaka:thermal}.
Then the analytic continuation of $ik_0\to\omega+i0^+$ leads to
\begin{align}\label{eq:response}
& R_{\C\C}(\omega+i0^+)
= \int_{\R\setminus\{0\}}\!\frac{d\nu}{2\pi i}\,f_{B,F}(\nu)\int_p \notag\\
& \times \bigl[G_+(\nu+\omega,p)G_+(\nu,p)\Gamma(\nu+\omega+i0^+,\nu+i0^+;p)^2 \notag\\
&\quad - G_+(\nu+\omega,p)G_-(\nu,p)\Gamma(\nu+\omega+i0^+,\nu-i0^+;p)^2 \notag\\
&\quad + G_+(\nu,p)G_-(\nu-\omega,p)\Gamma(\nu+i0^+,\nu-\omega-i0^+;p)^2 \notag\\
&\quad - G_-(\nu,p)G_-(\nu-\omega,p)\Gamma(\nu-i0^+,\nu-\omega-i0^+;p)^2\bigr],
\end{align}
where $f_{B,F}(\eps)=1/(e^{\beta\eps}\mp1)$ are the Bose-Einstein and Fermi-Dirac distribution functions and $G_\pm(\nu,p)\equiv G(\nu\pm i0^+,p)$ are the retarded and advanced propagators.
An important fact is that the product of retarded and advanced propagators with the same wave number has the singularity of
\begin{align}
G_+(\nu+\omega,p)G_-(\nu,p) = \frac{2\pi i\,\delta(\nu-\ep)}{\omega+i0^+} + O(\omega^0)
\end{align}
at zero frequency.
Consequently, Eq.~(\ref{eq:response}) substituted into Eq.~(\ref{eq:kubo}) gives rise to the Drude peak in Eq.~(\ref{eq:peak}) with its weight provided by
\begin{align}\label{eq:weight}
D = \beta\int_pf_{B,F}(\ep)\,[1\pm f_{B,F}(\ep)]\,[\Gamma_{+-}(p)]^2,
\end{align}
where $\Gamma_{+-}(p)\equiv\Gamma(\ep+i0^+,\ep-i0^+;p)$ is the on-shell vertex function.
Here we note that the two subtracted terms on the right-hand side of Eq.~(\ref{eq:stress}) have the role of imposing
\begin{subequations}\label{eq:sum-rule}
\begin{align}
& \int_pf_{B,F}(\ep)\,[1\pm f_{B,F}(\ep)]\,\Gamma_{+-}(p) = 0, \\
& \int_pf_{B,F}(\ep)\,[1\pm f_{B,F}(\ep)]\,\Gamma_{+-}(p)\,\ep = 0,
\end{align}
\end{subequations}
which are essential to define the static bulk viscosity in the Boltzmann equation~\cite{Dusling:2013,Chafin:2013,Fujii:preprint}.

It may be recalled that higher-order diagrams involving $n$ pairs of counterpropagating propagators have stronger singularities $\sim D\tau/(\omega\tau)^n$ at zero frequency and the resummation of them leads to finite transport coefficients of $D\tau$ in two and three dimensions~\cite{Enss:2011,Nishida:2019,Hofmann:2020,Fujii:2021,Fujii:preprint}.
However, one dimension is exceptional because the two-body interaction does not contribute to a finite relaxation time $\tau$ under the energy and momentum conservations.%
\footnote{On the other hand, the three-body interaction $\sim g_3$ in one dimension does contribute to the finite relaxation time $\tau\sim g_3^{-2}$ so as to reduce the Drude peak to $iD/(\omega+i/\tau)$~\cite{Tanaka:thermal}.}
More technically, such stronger singularities disappear due to cancellation among the self-energy and vertex (Maki-Thompson and Aslamazov-Larkin) corrections~\cite{Tanaka:thermal} (see the Appendix therein) so that the higher-order diagrams are to provide at most corrections to $D$.
Therefore, the Drude peak remains in one dimension and we now study its weights for the Lieb-Liniger model and the Cheon-Shigehara model.

\section{Lieb-Liniger model}\label{sec:bose}
The Hamiltonian density of the Lieb-Liniger model is provided by
\begin{align}
\hat\H = \frac{\d_x\hat\phi^\+\d_x\hat\phi}{2m}
+ \frac{g_B}{2}\hat\phi^\+\hat\phi^\+\hat\phi\hat\phi,
\end{align}
where $\hat\phi$ is the bosonic field operator and the coupling constant is related to the scattering length via $g_B=-2/ma>0$.
The stress operator can be found from the momentum continuity equation~\cite{Tanaka:bulk} so that the modified stress operator reads
\begin{align}
\hat\Pi = \frac{\hat\C}{ma} - \left(\frac{\d\C}{\d\N}\right)_\E\frac{\hat\N}{ma}
- \left(\frac{\d\C}{\d\E}\right)_\N\frac{\hat\H}{ma},
\end{align}
where $\hat\C=\hat\phi^\+\hat\phi^\+\hat\phi\hat\phi$ is the contact density operator with $\C=\<\hat\C\>$ and the total derivative of $\hat\N=\hat\phi^\+\hat\phi$ is suppressed because it vanishes under the spatial integration.

\begin{figure}[t]
(a)\hfill~\\
\includegraphics[width=\columnwidth]{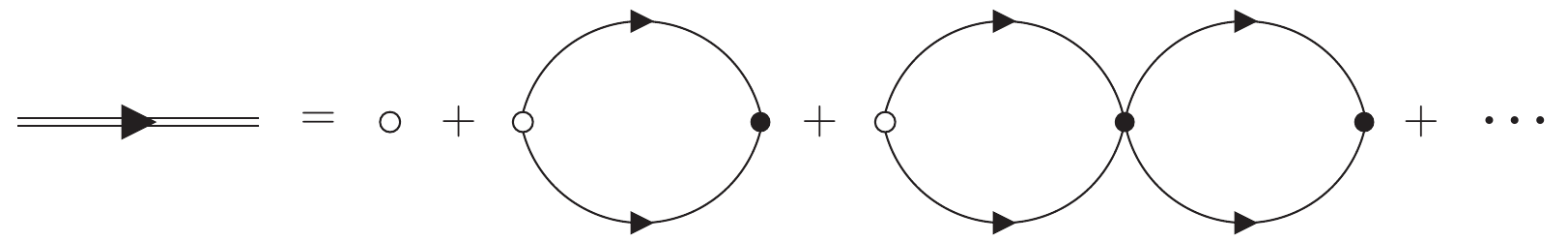}\\
(b)\hfill~\\
\includegraphics[width=0.5\columnwidth]{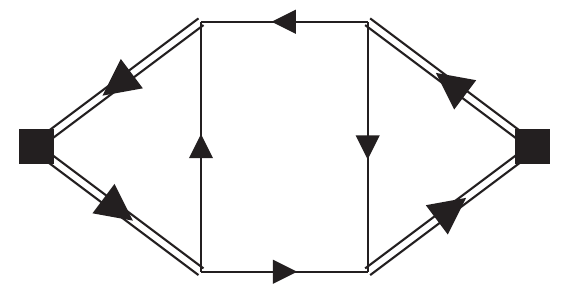}
\caption{\label{fig:contact}
Diagrammatic representation of (a) the pair propagators in Eqs.~(\ref{eq:propagator_bose}) and (\ref{eq:propagator_fermi}) and (b) the contact-contact response function contributing to the vertex functions in Eqs.~(\ref{eq:vertex_bose}) and (\ref{eq:vertex_fermi}).
The double line represents the pair propagator, whereas the closed circle is the interaction vertex carrying the coupling constant $-g_{B,F}/2$, the open circle carries just unity for bosons but $mg_F/2$ for fermions, and the closed square is to insert the contact density operator.}
\end{figure}

Although the static bulk viscosity of the Lieb-Liniger model was found to be finite both in the high-temperature limit and in the weak-coupling limit~\cite{Tanaka:bulk}, the Drude peak appears at higher orders from the diagram depicted in Fig.~\ref{fig:contact} for the contact-contact response function~\cite{Fujii:preprint}.
Together with other diagrams with single lines directly coupled to vertices provided by $\hat\N$ and $\hat\E$, it contributes to the stress-stress response function in Eq.~(\ref{eq:drude}) for the vertex function of
\begin{align}
& \Gamma(ik_0+ip_0,ip_0;p) = \frac{\gamma_B(ik_0+ip_0,ip_0;p)}{ma} \notag\\
&\quad - \left(\frac{\d\C}{\d\N}\right)_\E\frac1{ma}
- \left(\frac{\d\C}{\d\E}\right)_\N\frac{\eps_p}{ma}
\end{align}
with
\begin{align}\label{eq:vertex_bose}
& \gamma_B(ik_0+ip_0,ip_0;p)
= -\frac4\beta\sum_{q_0}\int_q\,G(iq_0,q) \notag\\
&\quad \times D_B(ip_0+iq_0,p+q)D_B(ik_0+ip_0+iq_0,p+q).
\end{align}
Here
\begin{align}\label{eq:propagator_bose}
& D_B(ip_0,p) = -\sum_{n=0}^\infty\left(-\frac{g_B}{2}\right)^n \notag\\
&\quad \times \left[\frac2\beta\sum_{q_0}\int_q\,G(ip_0-iq_0,p-q)G(iq_0,q)\right]^n
\end{align}
is the pair propagator in the medium and its diagrammatic representation is also depicted in Fig.~\ref{fig:contact}.%
\footnote{We cautiously note that $D_B(ip_0,p)$ and $D_F(ip_0,p)$ in this paper are slightly different from those in Ref.~\cite{Tanaka:bulk}.\label{footnote}}
The Matsubara frequency summation readily leads to
\begin{align}
\frac1{D_B(ip_0,p)} = -1
- \frac2{ma}\int_q\,\frac{1+f_B(\eps_{p/2-q})+f_B(\eps_{p/2+q})}
{ip_0-\eps_{p/2-q}-\eps_{p/2+q}}.
\end{align}

We first focus on the high-temperature limit at fixed number density, where the fugacity $z=e^{\beta\mu}\to0$ serves as a small parameter for the quantum virial expansion~\cite{Liu:2013}.
The vertex function in Eq.~(\ref{eq:vertex_bose}) to the lowest order in fugacity is evaluated as
\begin{align}\label{eq:vertex_high-T}
\gamma_B(p) = z\int_qe^{-\beta\eq}\frac{(p-q)^2}{\frac1{a^2}+\frac{(p-q)^2}{4}} + O(z^2)
\end{align}
for $\gamma_B(p)\equiv\gamma_B(\ep+i0^+,\ep-i0^+;p)$ under the analytic continuation of $ip_0\to\ep-i0^+$ followed by $ik_0\to i0^+$.
On the other hand, the contact density is provided by
\begin{align}
\C = \frac{2\sqrt2\,z^2}{\lambda_T}
\int_qe^{-2\beta\eq}\frac{q^2}{\frac1{a^2}+q^2} + O(z^3),
\end{align}
where $\lambda_T=\sqrt{2\pi\beta/m}$ is the thermal de Broglie wavelength~\cite{Nishida:2019,Tanaka:bulk}.
Its partial derivatives with respect to $\N$ and $\E$ are found to be
\begin{subequations}\label{eq:derivatives_high-T}
\begin{align}
\left(\frac{\d\C}{\d\N}\right)_\E &= \frac{\lambda_T}{z}
\left[3\C + \beta\left(\frac{\d\C}{\d\beta}\right)_{\beta\mu}\right] + O(z^2), \\
\left(\frac{\d\C}{\d\E}\right)_\N &= -\frac{2\beta\lambda_T}{z}
\left[\C + \beta\left(\frac{\d\C}{\d\beta}\right)_{\beta\mu}\right] + O(z^2),
\end{align}
\end{subequations}
respectively.
Therefore, $\Gamma_{+-}(p)=[\gamma_B(p)-(\d\C/\d\N)_\E-(\d\C/\d\E)_\N\,\ep]/ma$ resulting from Eqs.~(\ref{eq:vertex_high-T}) and (\ref{eq:derivatives_high-T}) is $O(z)$ and indeed satisfies
\begin{align}
\int_pe^{-\beta\ep}\,\Gamma_{+-}(p) = \int_pe^{-\beta\ep}\,\Gamma_{+-}(p)\,\ep = 0
\end{align}
in agreement with Eq.~(\ref{eq:sum-rule}).

\begin{figure}[t]
\includegraphics[width=0.9\columnwidth]{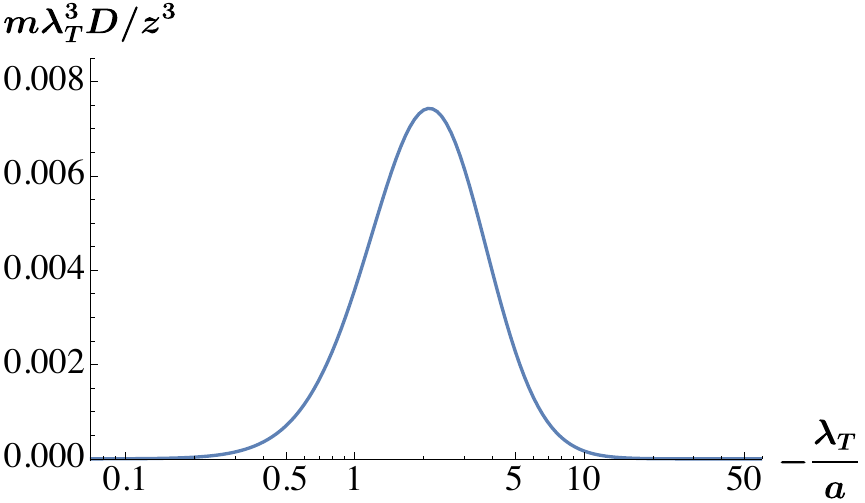}
\caption{\label{fig:drude_high-T}
Viscous Drude weight of the Lieb-Liniger model and the Cheon-Shigehara model in the high-temperature limit $z\to0$ from Eq.~(\ref{eq:drude_high-T}) as a function of the inverse scattering length.
The number density is provided by $\N=z/\lambda_T+O(z^2)$.}
\end{figure}

\begin{figure}[t]
(a)\hfill~\\
\includegraphics[width=0.9\columnwidth]{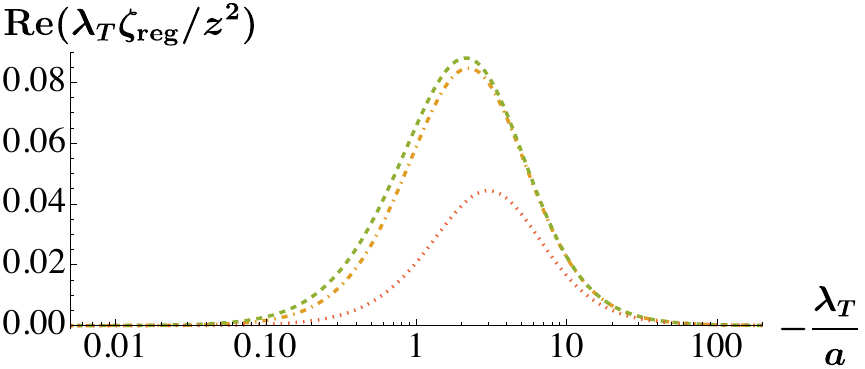}\\
(b)\hfill~\\
\includegraphics[width=0.9\columnwidth]{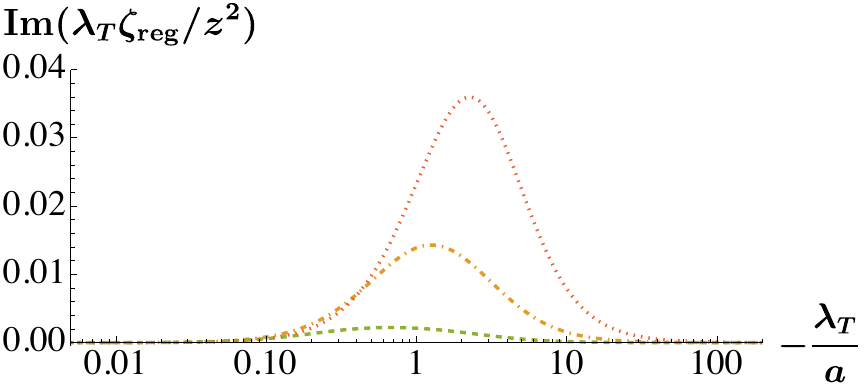}
\caption{\label{fig:regular_high-T}
Regular part of the frequency-dependent complex bulk viscosity for $z\to0$ from Eq.~(\ref{eq:regular_high-T}) as a function of the inverse scattering length.
Its (a) real and (b) imaginary parts are plotted for frequencies $\omega\,m\lambda_T^2=0.1$, 1, and 10 by green dashed, orange dot-dashed, and red dotted curves, respectively.}
\end{figure}

The Drude weight in the high-temperature limit is now obtained from Eq.~(\ref{eq:weight}) according to
\begin{align}\label{eq:drude_high-T}
D = z\beta\int_pe^{-\beta\ep}[\Gamma_{+-}(p)]^2 + O(z^4),
\end{align}
which is $O(z^3)$ to the lowest order in fugacity and varies nonmonotonically under the inverse scattering length as seen in Fig.~\ref{fig:drude_high-T}.
In particular, we find it to vanish as
\begin{align}\label{eq:high-T_weak-bose}
D \to \frac{16\sqrt3-27}{6\pi}\frac{z^3\lambda_T}{ma^4}
\end{align}
at $a\to-\infty$ and as
\begin{align}\label{eq:high-T_weak-fermi}
D \to 12\pi^3\frac{z^3a^6}{m\lambda_T^9}
\end{align}
at $a\to-0$.
For the sake of comparison, the regular part of the frequency-dependent complex bulk viscosity is $O(z^2)$~\cite{Tanaka:bulk},
\begin{align}\label{eq:regular_high-T}
& \zeta_\reg(\omega) = \frac{\sqrt2\,z^2}{(ma)^2\lambda_T}
\iint_{-\infty}^\infty\!\frac{d\eps}{\pi}\frac{d\eps'}{\pi}\,
\frac{e^{-\beta\eps} - e^{-\beta\eps'}}{\eps-\eps'} \notag\\
&\quad \times \frac{\Im[D_0(\eps-i0^+)]\Im[D_0(\eps'-i0^+)]}
{i(\omega+\eps-\eps'+i0^+)} + O(z^3),
\end{align}
which is plotted in Fig.~\ref{fig:regular_high-T} for several frequencies with $D_0(\eps)\equiv m/(1/a-\sqrt{-m\eps})$ being the pair propagator in the center-of-mass frame in the vacuum.

We then turn to the weak-coupling limit $a\to-\infty$, where the vertex function in Eq.~(\ref{eq:vertex_bose}) is evaluated as
\begin{align}\label{eq:vertex_weak-bose}
& \gamma_B(p) = 4\int_q\,f_B(\eq) \notag\\
&\quad - \frac{16}{ma}\int_{q,q'}\,
[f_B(\eps_{p+2q})-f_B(\eps_{p+q-q'}+\eps_{p+q+q'})] \notag\\
&\qquad \times \pv\frac{1+f_B(\eps_{p+q-q'})+f_B(\eps_{p+q+q'})}{\eq-\eqq} + O(a^{-2})
\end{align}
and the contact density is provided by
\begin{align}
\C = 2\int_{p,q}f_B(\ep)f_B(\eq) + O(a^{-1}).
\end{align}
Because its partial derivatives with respect to $\N$ and $\E$,
\begin{subequations}
\begin{align}
\left(\frac{\d\C}{\d\N}\right)_\E &= 4\int_qf_B(\eq) + O(a^{-1}), \\
\left(\frac{\d\C}{\d\E}\right)_\N &= O(a^{-1}),
\end{align}
\end{subequations}
are to cancel the first term on the right-hand side of Eq.~(\ref{eq:vertex_weak-bose}), the resulting $\Gamma_{+-}(p)=[\gamma_B(p)-(\d\C/\d\N)_\E-(\d\C/\d\E)_\N\,\ep]/ma$ is actually $O(a^{-2})$ to the lowest order in coupling.
Although $(\d\C/\d\N)_\E$ and $(\d\C/\d\E)_\N$ at $O(a^{-1})$ are directly computable, they are instead determined so as to satisfy Eq.~(\ref{eq:sum-rule}) for bosons.

\begin{figure}[t]
\includegraphics[width=0.9\columnwidth]{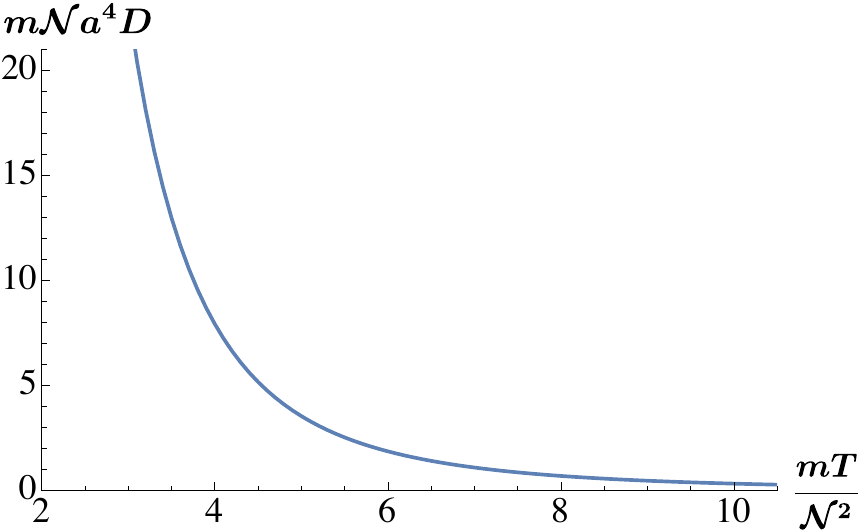}
\caption{\label{fig:drude_weak-bose}
Viscous Drude weight of the Lieb-Liniger model in the weak-coupling limit $a\to-\infty$, corresponding to the Cheon-Shigehara model in the strong-coupling limit, from Eq.~(\ref{eq:drude_weak-bose}) as a function of the temperature.
The number density is provided by $\N=\int_pf_B(\ep)+O(a^{-1})$.}
\end{figure}

\begin{figure}[t]
(a)\hfill~\\
\includegraphics[width=0.9\columnwidth]{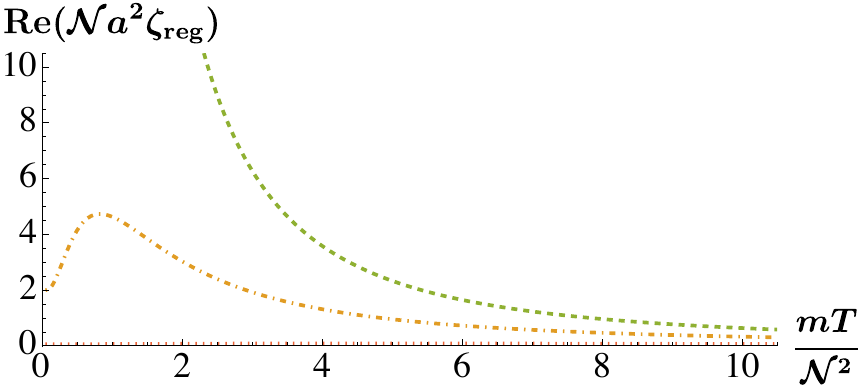}\\
(b)\hfill~\\
\includegraphics[width=0.9\columnwidth]{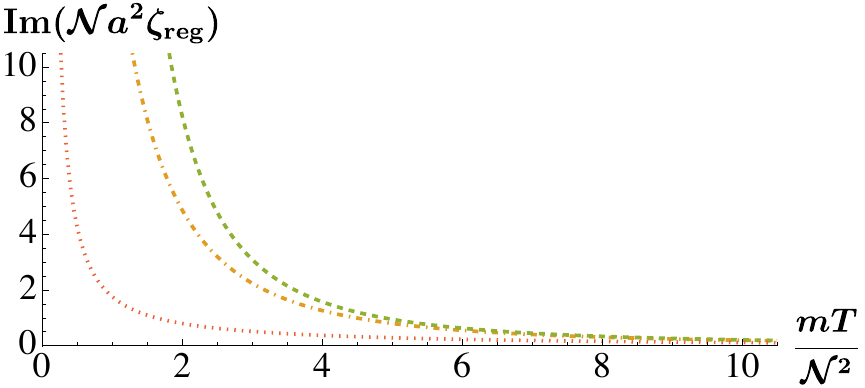}
\caption{\label{fig:regular_weak-bose}
Regular part of the frequency-dependent complex bulk viscosity for $a\to-\infty$ from Eq.~(\ref{eq:regular_weak-bose}) as a function of the temperature.
Its (a) real and (b) imaginary parts are plotted for frequencies $\omega\,m/\N^2=0.1$, 1, and 10 by green dashed, orange dot-dashed, and red dotted curves, respectively.}
\end{figure}

The Drude weight in the weak-coupling limit is now obtained from Eq.~(\ref{eq:weight}) according to
\begin{align}\label{eq:drude_weak-bose}
D = \beta\int_pf_B(\ep)\,[1+f_B(\ep)]\,[\Gamma_{+-}(p)]^2,
\end{align}
which is $O(a^{-4})$ to the lowest order in coupling and varies monotonically under the temperature as seen in Fig.~\ref{fig:drude_weak-bose} with its high-temperature limit being consistent with Eq.~(\ref{eq:high-T_weak-bose}).
We note that the subtractions in Eq.~(\ref{eq:stress}) are essential to correctly predict the order of $D$, which is otherwise misestimated at $O(a^{-2})$.
For the sake of comparison, the regular part of the frequency-dependent complex bulk viscosity is $O(a^{-2})$~\cite{Tanaka:bulk},
\begin{align}\label{eq:regular_weak-bose}
& \zeta_\reg(\omega) = \left(\frac2{ma}\right)^2\int_{p,q,q'}
\frac{f_B(\eps_{p/2-q}+\eps_{p/2+q}) - f_B(q\to q')}{2\eq-2\eqq} \notag\\
&\quad \times \frac{[1+f_B(\eps_{p/2-q})+f_B(\eps_{p/2+q})]
[\,q\to q'\,]}{i(\omega+2\eq-2\eqq+i0^+)} + O(a^{-3}),
\end{align}
which is plotted in Fig.~\ref{fig:regular_weak-bose} for several frequencies.
Thanks to the extended Bose-Fermi duality~\cite{Tanaka:bulk}, all the results described here are also applicable to the Cheon-Shigehara model in the strong-coupling limit.

\section{Cheon-Shigehara model}\label{sec:fermi}
The Hamiltonian density of the Cheon-Shigehara model is provided by
\begin{align}
\hat\H = \frac{\d_x\hat\psi^\+\d_x\hat\psi}{2m}
+ \frac{g_F}{2}(\d_x\hat\psi^\+)\hat\psi^\+\hat\psi(\d_x\hat\psi),
\end{align}
where $\hat\psi$ is the fermionic field operator and the coupling constant is related to the scattering length via
\begin{align}\label{eq:coupling_fermi}
\frac1{g_F} = -\frac{m\Lambda}{\pi} + \frac{m}{2a} < 0
\end{align}
with $\Lambda$ being the momentum cutoff for regularization~\cite{Cui:2016,Sekino:2021}.%
\footnote{See Ref.~\cite{Tanaka:bulk} for remarks regarding the three-body interaction term~\cite{Sekino:2018,Sekino:2021}, which is necessary for the complete correspondence to the Lieb-Liniger model but can be omitted for our analysis below within the two-body level.}
The stress operator can be found from the momentum continuity equation~\cite{Tanaka:bulk} so that the modified stress operator reads
\begin{align}
\hat\Pi = \frac{\hat\C}{ma} - \left(\frac{\d\C}{\d\N}\right)_\E\frac{\hat\N}{ma}
- \left(\frac{\d\C}{\d\E}\right)_\N\frac{\hat\H}{ma},
\end{align}
where $\hat\C=(mg_F/2)^2(\d_x\hat\psi^\+)\hat\psi^\+\hat\psi(\d_x\hat\psi)$ is the contact density operator with $\C=\<\hat\C\>$ and the total derivative of $\hat\N=\hat\psi^\+\hat\psi$ is suppressed because it vanishes under the spatial integration.

Although the static bulk viscosity of the Cheon-Shigehara model was found to be finite both in the high-temperature limit and in the weak-coupling limit~\cite{Tanaka:bulk}, the Drude peak appears at higher orders from the diagram depicted in Fig.~\ref{fig:contact} for the contact-contact response function~\cite{Fujii:preprint}.
Together with other diagrams with single lines directly coupled to vertices provided by $\hat\N$ and $\hat\E$, it contributes to the stress-stress response function in Eq.~(\ref{eq:drude}) for the vertex function of
\begin{align}
& \Gamma(ik_0+ip'_0,ip'_0;p) = \frac{\gamma_F(ik_0+ip'_0,ip'_0;p)}{ma} \notag\\
&\quad - \left(\frac{\d\C}{\d\N}\right)_\E\frac1{ma}
- \left(\frac{\d\C}{\d\E}\right)_\N\frac{\eps_p}{ma}
\end{align}
with
\begin{align}\label{eq:vertex_fermi}
& \gamma_F(ik_0+ip'_0,ip'_0;p)
= \frac1\beta\sum_{q'_0}\int_q\,(p-q)^2G(iq'_0,q) \notag\\
&\quad \times D_F(ip'_0+iq'_0,p+q)D_F(ik_0+ip'_0+iq'_0,p+q).
\end{align}
Here
\begin{align}\label{eq:propagator_fermi}
& D_F(ip_0,p) = -\frac{mg_F}{2}
\sum_{n=0}^\infty\left(-\frac{g_F}{2}\right)^n \notag\\
&\quad \times \left[\frac2\beta\sum_{q'_0}\int_q\,
(q-p/2)^2G(ip_0-iq'_0,p-q)G(iq'_0,q)\right]^n
\end{align}
is the pair propagator in the medium and its diagrammatic representation is also depicted in Fig.~\ref{fig:contact} (see footnote~\ref{footnote}).
The Matsubara frequency summation readily leads to
\begin{align}
& \frac1{D_F(ip_0,p)} = -\frac1a \notag\\
&\quad + 2\int_q\left[1 + \frac{q^2}{m}
\frac{1-f_F(\eps_{p/2-q})-f_F(\eps_{p/2+q})}{ip_0-\eps_{p/2-q}-\eps_{p/2+q}}\right],
\end{align}
where the regularization is applied under Eq.~(\ref{eq:coupling_fermi}) with $\Lambda\to\infty$.

We first focus on the high-temperature limit at fixed number density, where the vertex function in Eq.~(\ref{eq:vertex_fermi}) to the lowest order in fugacity is evaluated as
\begin{align}
\gamma_F(p) = z\int_qe^{-\beta\eq}\frac{(p-q)^2}{\frac1{a^2}+\frac{(p-q)^2}{4}} + O(z^2)
\end{align}
for $\gamma_F(p)\equiv\gamma_F(\ep+i0^+,\ep-i0^+;p)$ under the analytic continuation of $ip'_0\to\ep-i0^+$ followed by $ik_0\to i0^+$.
Therefore, the resulting expression proves to be identical to Eq.~(\ref{eq:vertex_high-T}) for the Lieb-Liniger model.
Because the equilibrium thermodynamic properties in Eq.~(\ref{eq:derivatives_high-T}) are also common~\cite{Girardeau:1960}, the Drude weight of the Cheon-Shigehara model in the high-temperature limit is provided by exactly the same formula as Eq.~(\ref{eq:drude_high-T}) to the lowest order in fugacity.
This is indeed expected and confirms the extended Bose-Fermi duality established by our recent work~\cite{Tanaka:bulk}.

We then turn to the weak-coupling limit $a\to-0$, where the vertex function in Eq.~(\ref{eq:vertex_fermi}) is evaluated as
\begin{align}\label{eq:vertex_weak-fermi}
& \gamma_F(p) = a^2\int_q\,(p-q)^2f_F(\eq)
+ \frac{16a^3}{m}\int_{q,q'}\,q^2q'^2 \notag\\
&\quad \times \biggl[\frac{f_F(\eps_{p+2q})}{\eqq}
+ [f_F(\eps_{p+2q})+f_B(\eps_{p+q-q'}+\eps_{p+q+q'})] \notag\\
&\qquad \times \pv\frac{1-f_F(\eps_{p+q-q'})-f_F(\eps_{p+q+q'})}{\eq-\eqq}\biggr] + O(a^4)
\end{align}
and the contact density is provided by
\begin{align}
\C = \frac{a^2}{2}\int_{p,q}(p-q)^2f_F(\ep)f_F(\eq) + O(a^3).
\end{align}
Because its partial derivatives with respect to $\N$ and $\E$,
\begin{subequations}
\begin{align}
\left(\frac{\d\C}{\d\N}\right)_\E &= a^2\int_qq^2f_F(\eq) + O(a^3), \\
\left(\frac{\d\C}{\d\E}\right)_\N &= 2ma^2\int_qf_F(\eq) + O(a^3),
\end{align}
\end{subequations}
are to cancel the first term on the right-hand side of Eq.~(\ref{eq:vertex_weak-fermi}), the resulting $\Gamma_{+-}(p)=[\gamma_F(p)-(\d\C/\d\N)_\E-(\d\C/\d\E)_\N\,\ep]/ma$ is apparently $O(a^2)$ to the lowest order in coupling.
Although $(\d\C/\d\N)_\E$ and $(\d\C/\d\E)_\N$ at $O(a^3)$ are directly computable, they are instead determined so as to satisfy Eq.~(\ref{eq:sum-rule}) for fermions.

\begin{figure}[t]
(a)\hfill~\\
\includegraphics[width=0.9\columnwidth]{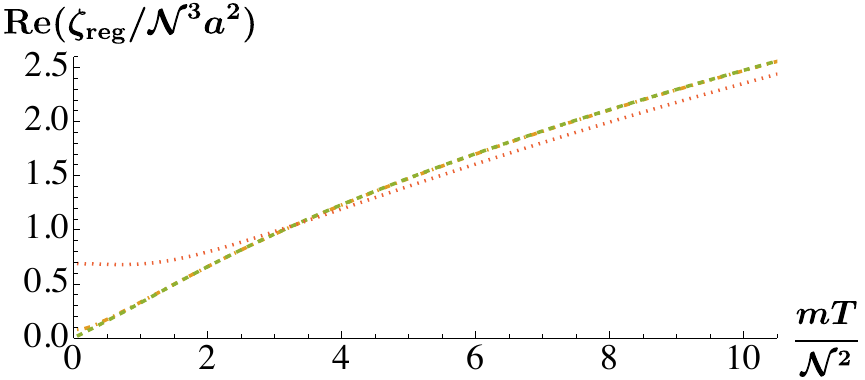}\\
(b)\hfill~\\
\includegraphics[width=0.9\columnwidth]{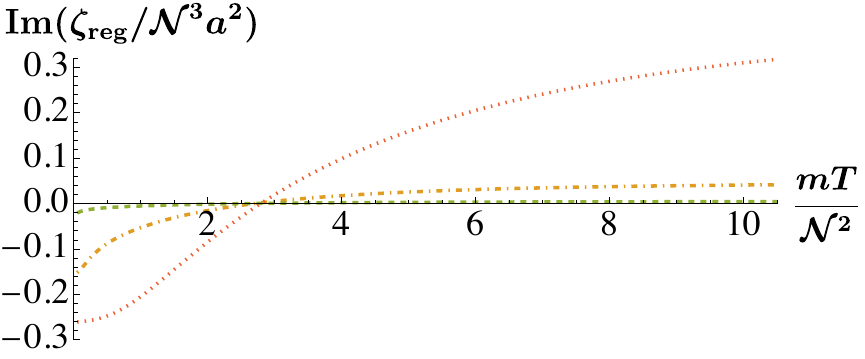}
\caption{\label{fig:regular_weak-fermi}
Regular part of the frequency-dependent complex bulk viscosity for $a\to-0$ from Eq.~(\ref{eq:regular_weak-fermi}) as a function of the temperature.
Its (a) real and (b) imaginary parts are plotted for frequencies $\omega\,m/\N^2=0.1$, 1, and 10 by green dashed, orange dot-dashed, and red dotted curves, respectively.
The number density is provided by $\N=\int_pf_F(\ep)+O(a)$.}
\end{figure}

The Drude weight in the weak-coupling limit is now obtained from Eq.~(\ref{eq:weight}) according to
\begin{align}\label{eq:drude_weak-fermi}
D = \beta\int_pf_F(\ep)\,[1-f_F(\ep)]\,[\Gamma_{+-}(p)]^2,
\end{align}
which is however found to vanish numerically at $O(a^4)$.%
\footnote{We also confirmed numerically that Eq.~(\ref{eq:vertex_weak-fermi}) is actually reduced to $\gamma_F(p)|_{O(a^3)}=3a^3\int_{q,q'}(p^2+2q^2)f(\eq)f(\eqq)$ so as to make $\Gamma_{+-}(p)=O(a^3)$.}
Therefore, the nonvanishing Drude weight is expected to appear at $O(a^6)$ to the lowest order in coupling as also indicated by Eq.~(\ref{eq:high-T_weak-fermi}) in the high-temperature limit.
We note that the subtractions in Eq.~(\ref{eq:stress}) are essential to correctly predict the order of $D$, which is otherwise misestimated at $O(a^2)$.
For the sake of comparison, the regular part of the frequency-dependent complex bulk viscosity is $O(a^2)$~\cite{Tanaka:bulk},
\begin{align}\label{eq:regular_weak-fermi}
& \zeta_\reg(\omega) = \left(\frac{2a}{m}\right)^2\int_{p,q,q'}
\frac{f_B(\eps_{p/2-q}+\eps_{p/2+q}) - f_B(q\to q')}{2\eq-2\eqq} \notag\\
&\quad \times q^2q'^2\frac{[1-f_F(\eps_{p/2-q})-f_F(\eps_{p/2+q})]
[\,q\to q'\,]}{i(\omega+2\eq-2\eqq+i0^+)} + O(a^3),
\end{align}
which is plotted in Fig.~\ref{fig:regular_weak-fermi} for several frequencies.
Thanks to the extended Bose-Fermi duality~\cite{Tanaka:bulk}, all the results described here are also applicable to the Lieb-Liniger model in the strong-coupling limit.

\section{Summary}
In summary, we showed that the frequency-dependent complex bulk viscosities of one-dimensional Bose and Fermi gases with contact interactions have the structure of Eq.~(\ref{eq:peak}), which is contributed to by the Drude peak divergent at zero frequency.
Essential for the Drude peak to remain in one dimension is the fact that the two-body interaction does not contribute to a finite relaxation time under the energy and momentum conservations unlike in higher dimensions.
In particular, the Drude weights were evaluated based on the Kubo formula in the high-temperature, weak-coupling, and strong-coupling limits (shaded regions in Fig.~\ref{fig:duality}), where systematic expansions in terms of small parameters are available.
In all three limits, the Drude peaks appear at higher orders compared to the finite regular parts so that
\begin{align}
\zeta_\reg(\omega) \sim O(z^2), \quad D \sim O(z^3)
\end{align}
for $z\to0$,
\begin{align}
\zeta_\reg(\omega) \sim O(a^{-2}), \quad D \sim O(a^{-4})
\end{align}
for $a\to-\infty$, and
\begin{align}
\zeta_\reg(\omega) \sim O(a^2), \quad D \sim O(a^6)
\end{align}
for $a\to-0$ as quantitatively determined in Figs.~\ref{fig:drude_high-T}--\ref{fig:regular_weak-fermi}.
Our findings are consistent with the divergent bulk viscosity at zero frequency for integrable systems~\cite{Matveev:2017} as well as with the vanishing bulk viscosity at any frequency for conformal systems~\cite{Son:2007,Taylor:2010}, but do not corroborate the argument in Ref.~\cite{Maki:preprint} for finite bulk viscosity in spite of integrability.

Finally, we note that the frequency-dependent complex bulk viscosity can be extracted experimentally with ultracold atoms by modulating the scattering length periodically as $a(t)=a+\delta a\sin(\omega t)$.
Here the linear-response theory predicts that the contact density responds according to
\begin{align}
\C(t) - \C_\mathrm{eq}[a(t)]
&= \Im\left[\frac{R_{\C\C}(\omega+i0^+)-R_{\C\C}(0)}{ma^2}\,
\delta a\,e^{-i\omega t}\right] \notag\\
&\quad + O(\delta a^2),
\end{align}
as well as the energy density and the entropy density produced at the rates of
\begin{align}
\dot\E(t) &= \frac{\C(t)}{ma^2(t)}\,\dot{a}(t), \\
T\dot\S(t) &= \frac{\C(t)-\C_\mathrm{eq}[a(t)]}{ma^2(t)}\,\dot{a}(t),
\end{align}
with $\C_\mathrm{eq}[a]$ being the equilibrium contact density for the scattering length $a$~\cite{Fujii:2018,Nishida:2019}.
Therefore, by measuring the contact, energy, or entropy density under the periodic modulation of the scattering length, it is possible to extract the contact-contact response function and thus the frequency-dependent complex bulk viscosity via Eq.~(\ref{eq:kubo}) with $R_{\Pi\Pi}(w)=R_{\C\C}(w)/(ma)^2$.
In particular, the hydrodynamic entropy production rate is obtained at low frequency $\omega\to0$,
\begin{align}
T\dot\S(t) = \underbrace{\frac{\Im[R_{\Pi\Pi}(\omega+i0^+)]}{\omega}}_{\to\,\zeta}
\left[\frac{\dot{a}(t)}{a(t)}\right]^2 + O(\omega^4,\delta a^3),
\end{align}
where $-\dot{a}(t)/a(t)$ is to simulate the fluid expansion rate~\cite{Fujii:2018}.

\acknowledgments
The author thanks Keisuke Fujii and Tomohiro Tanaka for valuable discussions.
This work was supported by JSPS KAKENHI Grants No.\ JP18H05405 and No.\ JP21K03384.


\begin{thebibliography}{99}

\bibitem{Kohn:1964}
W.~Kohn,
``Theory of the insulating state,''
\href{https://doi.org/10.1103/PhysRev.133.A171}
{Phys.\ Rev.\ \textbf{133}, A171-A181 (1964)}.

\bibitem{Mahan}
See, for example, G.~D.~Mahan,
\textit{Many-Particle Physics,} 3rd ed.\
(Kluwer Academic/Plenum, New York, 2000).

\bibitem{Castella:1995}
H.~Castella, X.~Zotos, and P.~Prelov\v{s}ek,
``Integrability and ideal conductance at finite temperatures,''
\href{https://doi.org/10.1103/PhysRevLett.74.972}
{Phys.\ Rev.\ Lett.\ \textbf{74}, 972-975 (1995)}.

\bibitem{Zotos:1997}
X.~Zotos, F.~Naef, and P.~Prelov\v{s}ek,
``Transport and conservation laws,''
\href{https://doi.org/10.1103/PhysRevB.55.11029}
{Phys.\ Rev.\ B \textbf{55}, 11029-11032 (1997)}.

\bibitem{Zotos:2004}
X.~Zotos and P. Prelov\v{s}ek,
``Transport in one dimensional quantum systems,''
\href{https://doi.org/10.1007/978-1-4020-3463-3_11}
{in \textit{Strong Interactions in Low Dimensions,} edited by D.~Baeriswyl and L.~Degiorgi, Physics and Chemistry of Materials with Low-Dimensional Structures Vol.~25 (Springer, Dordrecht, 2004), Chap.~11}.

\bibitem{Zotos:2005}
X.~Zotos,
``Issues on the transport of one dimensional quantum systems,''
\href{https://doi.org/10.1143/JPSJS.74S.173}
{J.\ Phys.\ Soc.\ Jpn.\ \textbf{74}, 173-180 (2005)}.

\bibitem{Sirker:2020}
J.~Sirker,
``Transport in one-dimensional integrable quantum systems,''
\href{https://doi.org/10.21468/SciPostPhysLectNotes.17}
{SciPost Phys.\ Lect.\ Notes \textbf{17}, 1-19 (2020)}.

\bibitem{Bertini:2021}
B.~Bertini, F.~Heidrich-Meisner, C.~Karrasch, T.~Prosen, R.~Steinigeweg, and M.~\v{Z}nidari\v{c},
``Finite-temperature transport in one-dimensional quantum lattice models,''
\href{https://doi.org/10.1103/RevModPhys.93.025003}
{Rev.\ Mod.\ Phys.\ \textbf{93}, 025003 (2021)}.

\bibitem{Matveev:2017}
K.~A.~Matveev and M.~Pustilnik,
``Viscous dissipation in one-dimensional quantum liquids,''
\href{https://doi.org/10.1103/PhysRevLett.119.036801}
{Phys.\ Rev.\ Lett.\ \textbf{119}, 036801 (2017)}.

\bibitem{DeGottardi:2020}
W.~DeGottardi and K.~A.~Matveev,
``Viscous properties of a degenerate one-dimensional Fermi gas,''
\href{https://doi.org/10.1103/PhysRevLett.125.076601}
{Phys.\ Rev.\ Lett.\ \textbf{125}, 076601 (2020)}.

\bibitem{Tanaka:bulk}
T.~Tanaka and Y.~Nishida,
``Bulk viscosity of dual Bose and Fermi gases in one dimension,''
\href{https://doi.org/10.1103/PhysRevLett.129.200402}
{Phys.\ Rev.\ Lett.\ \textbf{129}, 200402 (2022)}.

\bibitem{Lieb:1963a}
E.~H.~Lieb and W.~Liniger,
``Exact analysis of an interacting Bose gas. I. The general solution and the ground state,''
\href{https://doi.org/10.1103/PhysRev.130.1605}
{Phys.\ Rev.\ \textbf{130}, 1605-1616 (1963)}.

\bibitem{Lieb:1963b}
E.~H.~Lieb,
``Exact analysis of an interacting Bose gas. II. The excitation spectrum,''
\href{https://doi.org/10.1103/PhysRev.130.1616}
{Phys.\ Rev.\ \textbf{130}, 1616-1624 (1963)}.

\bibitem{Cheon:1999}
T.~Cheon and T.~Shigehara,
``Fermion-boson duality of one-dimensional quantum particles with generalized contact interactions,''
\href{https://doi.org/10.1103/PhysRevLett.82.2536}
{Phys.\ Rev.\ Lett.\ \textbf{82}, 2536-2539 (1999)}.

\bibitem{Fujii:2020}
K.~Fujii and Y.~Nishida,
``Bulk viscosity of resonating fermions revisited: Kubo formula, sum rule, and the dimer and high-temperature limits,''
\href{https://doi.org/10.1103/PhysRevA.102.023310}
{Phys.\ Rev.\ A \textbf{102}, 023310 (2020)}.%
\footnote{There is a typo in Ref.~\cite{Fujii:2020}: $\Im[m^2D(E-i0^+,\bm k+\bm p)]\Im[m^2D(E'-i0^+,\bm p)]$ in Eq.~(51) should be $\Im[m^2D(E'-i0^+,\bm k+\bm p)]\Im[m^2D(E-i0^+,\bm p)]$.}

\bibitem{Mori:1962}
H.~Mori,
``Collective motion of particles at finite temperatures,''
\href{https://doi.org/10.1143/PTP.28.763}
{Prog.\ Theor.\ Phys.\ \textbf{28}, 763-783 (1962)}.

\bibitem{Luttinger:1964}
J.~M.~Luttinger,
``Theory of thermal transport coefficients,''
\href{https://doi.org/10.1103/PhysRev.135.A1505}
{Phys.\ Rev.\ \textbf{135}, A1505-A1514 (1964)}.

\bibitem{Bradlyn:2012}
B.~Bradlyn, M.~Goldstein, and N.~Read,
``Kubo formulas for viscosity: Hall viscosity, Ward identities, and the relation with conductivity,''
\href{https://doi.org/10.1103/PhysRevB.86.245309}
{Phys.\ Rev.\ B \textbf{86}, 245309 (2012)}.

\bibitem{Fujii:2021}
K.~Fujii and Y.~Nishida,
``Microscopic derivation of the Boltzmann equation for transport coefficients of resonating fermions at high temperature,''
\href{https://doi.org/10.1103/PhysRevA.103.053320}
{Phys.\ Rev.\ A \textbf{103}, 053320 (2021)}.

\bibitem{Tanaka:thermal}
T.~Tanaka and Y.~Nishida,
``Thermal conductivity of a weakly interacting Bose gas in quasi-one-dimension,''
\href{https://doi.org/10.1103/PhysRevE.106.064104}
{Phys.\ Rev.\ E \textbf{106}, 064104 (2022)}.

\bibitem{Dusling:2013}
K.~Dusling and T.~Sch\"afer,
``Bulk viscosity and conformal symmetry breaking in the dilute Fermi gas near unitarity,''
\href{https://doi.org/10.1103/PhysRevLett.111.120603}
{Phys.\ Rev.\ Lett.\ \textbf{111}, 120603 (2013)}.

\bibitem{Chafin:2013}
C.~Chafin and T.~Sch\"afer,
``Scale breaking and fluid dynamics in a dilute two-dimensional Fermi gas,''
\href{https://doi.org/10.1103/PhysRevA.88.043636}
{Phys.\ Rev.\ A \textbf{88}, 043636 (2013)}.

\bibitem{Fujii:preprint}
K.~Fujii and T.~Enss,
``Bulk viscosity of resonantly interacting fermions in the quantum virial expansion,''
\href{https://doi.org/10.48550/arXiv.2208.03353}
{arXiv:2208.03353 [cond-mat.quant-gas]}.

\bibitem{Enss:2011}
T.~Enss, R.~Haussmann, and W.~Zwerger,
``Viscosity and scale invariance in the unitary Fermi gas,''
\href{https://doi.org/10.1016/j.aop.2010.10.002}
{Ann.\ Phys.\ (NY) \textbf{326}, 770-796 (2011)}.

\bibitem{Nishida:2019}
Y.~Nishida,
``Viscosity spectral functions of resonating fermions in the quantum virial expansion,''
\href{https://doi.org/10.1016/j.aop.2019.167949}
{Ann.\ Phys.\ (NY) \textbf{410}, 167949 (2019)}.%
\footnote{There is a typo in Ref.~\cite{Nishida:2019}: $\eps$ on both sides of Eq.~(73) should be $\omega$.}

\bibitem{Hofmann:2020}
J.~Hofmann,
``High-temperature expansion of the viscosity in interacting quantum gases,''
\href{https://doi.org/10.1103/PhysRevA.101.013620}
{Phys.\ Rev.\ A \textbf{101}, 013620 (2020)}.

\bibitem{Liu:2013}
X.-J.~Liu,
``Virial expansion for a strongly correlated Fermi system and its application to ultracold atomic Fermi gases,''
\href{https://doi.org/10.1016/j.physrep.2012.10.004}
{Phys.\ Rep.\ \textbf{524}, 37-83 (2013)}.

\bibitem{Cui:2016}
X.~Cui,
``Universal one-dimensional atomic gases near odd-wave resonance,''
\href{https://doi.org/10.1103/PhysRevA.94.043636}
{Phys.\ Rev.\ A \textbf{94}, 043636 (2016)}.

\bibitem{Sekino:2021}
Y.~Sekino and Y.~Nishida,
``Field-theoretical aspects of one-dimensional Bose and Fermi gases with contact interactions,''
\href{https://doi.org/10.1103/PhysRevA.103.043307}
{Phys.\ Rev.\ A \textbf{103}, 043307 (2021)}.

\bibitem{Sekino:2018}
Y.~Sekino, S.~Tan, and Y.~Nishida,
``Comparative study of one-dimensional Bose and Fermi gases with contact interactions from the viewpoint of universal relations for correlation functions,''
\href{https://doi.org/10.1103/PhysRevA.97.013621}
{Phys.\ Rev.\ A \textbf{97}, 013621 (2018)}.

\bibitem{Girardeau:1960}
M.~Girardeau,
``Relationship between systems of impenetrable bosons and fermions in one dimension,''
\href{https://doi.org/10.1063/1.1703687}
{J.\ Math.\ Phys.\ \textbf{1}, 516-523 (1960)}.

\bibitem{Son:2007}
D.~T.~Son,
``Vanishing bulk viscosities and conformal invariance of the unitary Fermi gas,''
\href{https://doi.org/10.1103/PhysRevLett.98.020604}
{Phys.\ Rev.\ Lett.\ \textbf{98}, 020604 (2007)}.

\bibitem{Taylor:2010}
E.~Taylor and M.~Randeria,
``Viscosity of strongly interacting quantum fluids: Spectral functions and sum rules,''
\href{https://doi.org/10.1103/PhysRevA.81.053610}
{Phys.\ Rev.\ A \textbf{81}, 053610 (2010)}.

\bibitem{Maki:preprint}
J.~Maki and S.~Zhang,
``Viscous flow in a 1D spin-polarized Fermi gas: The role of integrability on viscosity,''
\href{https://doi.org/10.48550/arXiv.2206.08109}
{arXiv:2206.08109 [cond-mat.quant-gas]}.

\bibitem{Fujii:2018}
K.~Fujii and Y.~Nishida,
``Hydrodynamics with spacetime-dependent scattering length,''
\href{https://doi.org/10.1103/PhysRevA.98.063634}
{Phys.\ Rev.\ A \textbf{98}, 063634 (2018)}.

\end{thebibliography}
\end{document}